\documentclass[aps,prl,twocolumn,showpacs,superscriptaddress]{revtex4-1}

\usepackage{amsfonts}
\usepackage{amsmath}
\usepackage{amssymb}
\usepackage{graphicx}
\usepackage{bm}

\begin{document}


\title{In-plane anisotropy of electrical resistivity in the strain-detwinned SrFe$_2$As$_2$}

\author{E.~C.~Blomberg}

\affiliation{Ames Laboratory, Ames, Iowa 50011, USA}
\affiliation{Department of Physics and Astronomy, Iowa State University, Ames,
Iowa 50011, USA }

\author{M.~A.~Tanatar}

\email[Corresponding author: ]{tanatar@ameslab.gov}

\affiliation{Ames Laboratory, Ames, Iowa 50011, USA}

\author{A.~Kreyssig}

\affiliation{Ames Laboratory, Ames, Iowa 50011, USA}
\affiliation{Department of Physics and Astronomy, Iowa State University, Ames,
Iowa 50011, USA }

\author{N.~Ni}

\affiliation{Ames Laboratory, Ames, Iowa 50011, USA}

\affiliation{Department of Physics and Astronomy, Iowa State University, Ames,
Iowa 50011, USA }

\author{A.~Thaler}

\affiliation{Ames Laboratory, Ames, Iowa 50011, USA}

\affiliation{Department of Physics and Astronomy, Iowa State University, Ames,
Iowa 50011, USA }

\author{Rongwei~Hu}

\affiliation{Department of Physics and Astronomy, Iowa State University, Ames,
Iowa 50011, USA }

\author{S.~L.~Bud'ko}

\affiliation{Ames Laboratory, Ames, Iowa 50011, USA}

\affiliation{Department of Physics and Astronomy, Iowa State University, Ames,
Iowa 50011, USA }

\author{P.~C.~Canfield}

\affiliation{Ames Laboratory, Ames, Iowa 50011, USA}

\affiliation{Department of Physics and Astronomy, Iowa State University, Ames,
Iowa 50011, USA }

\author{A.~I.~Goldman}

\affiliation{Ames Laboratory, Ames, Iowa 50011, USA}

\affiliation{Department of Physics and Astronomy, Iowa State University, Ames,
Iowa 50011, USA }

\author{R.~Prozorov}

\affiliation{Ames Laboratory, Ames, Iowa 50011, USA}

\affiliation{Department of Physics and Astronomy, Iowa State University, Ames,
Iowa 50011, USA }

\date{31 December 2010}

\begin{abstract}
Intrinsic, in-plane anisotropy of electrical resistivity was studied on mechanically detwinned single crystals of SrFe$_2$As$_2$ above and below the temperature of the coupled structural/magnetic transition, $T_{\textrm{TO}}$. Resistivity is smaller for electrical current flow along the orthorhombic $a_o$ direction (direction of antiferromagnetically alternating magnetic moments) and is larger for transport along the $b_o$ direction (direction of ferromagnetic chains), which is similar to CaFe$_2$As$_2$ and BaFe$_2$As$_2$ compounds. A strongly first order structural transition in SrFe$_2$As$_2$ was confirmed by high-energy x-ray measurements, with the transition temperature, and character unaffected by moderate strain. For small strain levels, which are just sufficient to detwin the sample, we find a negligible effect on the resistivity above $T_{\textrm{TO}}$.  With the increase of strain, the resistivity anisotropy starts to develop above $T_{\textrm{TO}}$, clearly showing the relation of anisotropy to an anomalously strong response to strain. Our study suggests that electronic nematicity cannot be observed in the FeAs based compounds in which the structural transition is strongly first order.

\end{abstract}


\maketitle



\section{Introduction}

Parent compounds of high transition temperature iron arsenide superconductors, AFe$_{2}$As$_{2}$ (A =  alkali earth Ca, Sr, Ba) \cite{Rotter,Yan,Ca-phasetransition,Milton}, crystallize in a tetragonal symmetry lattice. On cooling below a characteristic temperature $T_{\textrm{TO}}$ (135 K for A=Ba, 170 K for A=Ca and $\sim$205~K for A=Sr) the crystals undergo a tetragonal-to-orthorhombic structural phase transition (and magnetic transition to an antiferromagnetic state). Free - standing samples develop four crystallographically equivalent domains \cite{domains}, and their random distribution in the crystals masks any intrinsic in-plane anisotropy of the orthorhombic phase in bulk measurements, such as electrical resistivity. Recent studies of anisotropy on detwinned crystals of the parent compounds of iron arsenide superconductors CaFe$_2$As$_2$ and BaFe$_2$As$_2$ (Ca122 and Ba122 in the following)  \cite{detwinning}, of Co-doped Ba(Fe$_{1-x}$Co$_x$)$_2$As$_2$, BaCo122 \cite{Fisher}, and of EuFe$_{2-x}$Co$_x$As$_2$, EuCo122, \cite{Eu} found that there is a subtle difference in the character of the structural/magnetic ordering between the compounds. Ca122 shows pronounced in plane electronic anisotropy of the orthorhombic (O) phase, with resistivity along the orthorhombic $a_o$ axis, $\rho_{ao}$, becoming smaller than that along the $b_o$ axis, $\rho_{bo}$, in the whole temperature range $T< T_{\textrm{TO}}$. This anisotropy vanishes in the crystallographically isotropic tetragonal (T) phase above the temperature of the coupled structural/magnetic, strongly first order, transition. Similarly, in Ba122 and especially BaCo122 and EuCo122, the anisotropy is found in the orthorhombic phase below $T_{\textrm{TO}}$, but unlike in Ca122, it does not vanish immediately at $T_{\textrm{TO}}$. This difference was related to the difference in the type of the phase transition: strong first order transition in Ca122 \cite{Ca-phasetransition}, almost second order in parent Ba122 \cite{RotterSDW} and clearly second order, with split structural and magnetic transitions, in doped BaCo122 compounds \cite{NiNiCo}.

The anisotropy in BaCo122 for $T> T_{\textrm{TO}}$ was ascribed to formation of an electronic nematic phase \cite{Fisher,detwinning}, a "translationally invariant metallic phase with a spontaneously generated spatial anisotropy" \cite{Fradkin}, as originally suggested theoretically for pnictides  \cite{Johannes}. A similar phase with intrinsic, in plane, electronic anisotropy in the high symmetry tetragonal phase is also found in Sr$_3$Ru$_2$O$_7$ \cite{327} and quantum Hall effect systems \cite{quantumHall}. Additional two-fold electronic anisotropy in the orthorhombic phase is found in the  high temperature cuprate superconductors \cite{nematiccuprates}. Several theoretical models for the explanation of this phase were suggested, for recent review see \cite{Fradkin}. Because of its proximity to superconductivity, this phase is of great interest. On the other hand, a less exotic possibility is that the strain vector breaks rotational invariance and, in these very pressure-sensitive materials, gives rise to anisotropy. Alternatively, the anisotropy can be induced by structural precursor effects \cite{Vaknin}. In this situation, it is of extreme importance to understand the connection between the electronic anisotropy above the transition and the phase with structural orthorhombic distortion below $T_{\textrm{TO}}$.

In this paper we study electronic anisotropy of the third member of the 122 family of parent compounds, SrFe$_2$As$_2$  (Sr122) in which we demonstrate nearly complete mechanical detwinning of single crystals through the application of uniaxial mechanical strain. The samples show a clear first-order structural transition as directly observed by synchrotron x-ray measurements. Furthermore, in sharp contrast with Ba122 and BaCo122, we find no anisotropy in the tetragonal crystallographic phase of Sr122, which is in line with theoretical predictions \cite{Fradkin}.
The anisotropy of resistivity can, however, be induced above $T_{\textrm{TO}}$ by applying a mild mechanical strain showing extreme sensitivity of the compounds to uniaxial strain. Our results suggest that directly associating electronic anisotropy in a tetragonal phase with nematicity is not trivial, and requires independent verification of the effect of the strain.

\section{Experimental}

\begin{figure}[tb]
\includegraphics[width=0.8\linewidth]{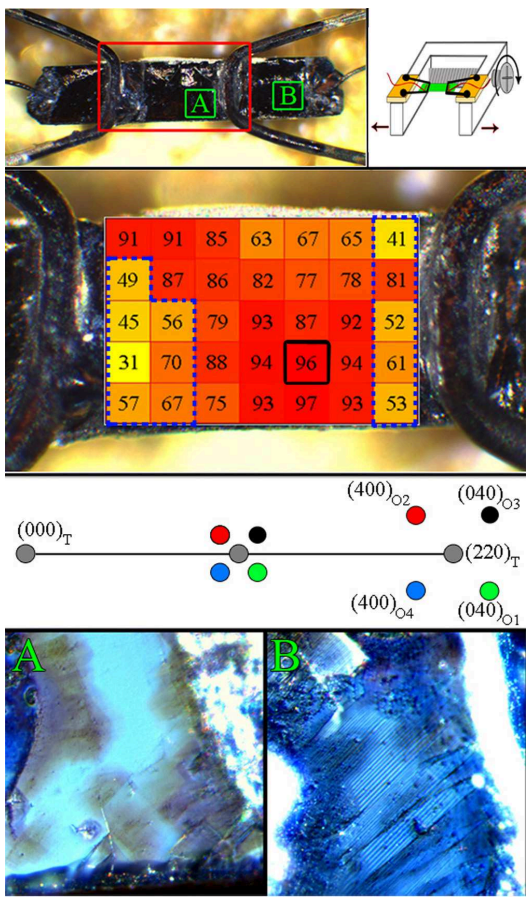}
\caption{(Color online) Top panel- a schematic of the sample mounted on a ``horseshoe'' with strain applied through the potential wires by adjusting a push-screw. The wires were insulated from the horseshoe by thin fiberglass boards. The second panel shows a zoom of the central area of the sample, overlaid with a spatial map, taken at 6~K, of the percentage of domain population with orthorhombic distortion along the strain (domains O2, O4 in the schematic presentation of the x-ray Laue pattern, third panel) as determined from the integrated x-ray intensity over all four possible domains, with actual x-ray data shown below in Fig.~\ref{peaks}. Thick dashed lines at potential contacts show the area above the soldered contact. The bottom panel shows polarized optical microscopy images of the strained (left) and unstrained (right)  areas at 5~K, revealing mechanical detwinning on the surface  of the sample between potential contacts. }
\label{detwinning}
\end{figure}

Single crystals of Sr122 were grown out of tin flux and were characterized by single crystal x-ray diffraction \cite{Yan}. Resistivity measurements were also reproduced on FeAs flux grown \cite{Ca-phasetransition,Yan} and annealed \cite{Suchitra} crystals, showing a similar transition temperature to Sn-grown samples.
The crystals were cut into strips along the tetragonal [110] $_T$ direction (which below $T_{\textrm{TO}}$ becomes either the [100]$_o$ a-axis or [010]$_o$ b-axis in the orthorhombic phase.) Typically samples had dimensions $(2-3) \times 0.5 \times (0.05-0.1)$ mm$^3$. Mechanical strain was applied through either thick (0.125~mm) or thin (0.05~mm) silver wires, soldered \cite{contacts_SUST} to form potential probes \cite{detwinning}, see top panel in  Fig.~\ref{detwinning}. The ends of the
wires were mounted on two insulator boards attached to a brass horseshoe. The horseshoe was deformed by a stainless push-screw, and thereby strained the crystals. Thin silver wires (~0.050 mm) were soldered to the ends of the samples to form current leads. These wires were bent so as to create minimal strain. Use of thinner wires for transmission of strain than in our initial study \cite{detwinning} resulted in improved control of strain in the samples. Four-probe resistance measurements were carried out in a {\it Quantum Design PPMS} from 5~K to 300~K. Visualization of structural domains in unstrained samples and their absence in detwinned samples was performed in a $^4$He flow-type cryostat mounted on the table of a polarized - light \textit{Leica DMLM} microscope Ref.~\onlinecite{domains}. Samples were imaged before and after the application of strain from room temperature to 5 K. The highest contrast of images was achieved for a configuration when the tetragonal [100] direction was 45$^o$ with respect to the polarization plane.

High-energy x-ray measurements of detwinned Sr122 were made at the MU-CAT sector (beamline 6ID-6) of the Advanced Photon Source at Argonne National Lab. Measurements were made on a sample grown out of Sn-flux (sample \#1) selected from 5 resistively and optically characterized samples by the criterion of the sharpness of the resistive features. Measurements using high-energy x-rays were made from
6K to 215K in 10K increments through the entire temperature range and at 1K increments in the vicinity of the structural transition. Entire reciprocal planes were recorded using the method described in detail in Ref.\onlinecite{Kreyssig07} which has been successfully
applied recently to study the domain structure in pnictides \cite{domains,domains2,detwinning}.
The absorption length of the high energy (99.3\,keV) x-rays was about
1.5~mm. This allowed for full penetration through the typically 0.05 to 0.1~mm thick samples, mounted with their \textbf{c} direction parallel to the incident $x$-ray beam. The beam size was reduced to $0.2 \times 0.2$ mm$^{2}$ by a slit system. As a result, each measurement averages over the entire
sample \textit{volume} selected by the beam dimension in the (\textbf{ab}) plane and its projection through the sample along the \textbf{c} direction. The direct beam was blocked by a beam stop behind the sample. Two-dimensional scattering patterns were measured by a MAR345 image-plate positioned
1730\,mm behind the sample. During the measurement, the sample was tilted through two independent angles, $\mu$ and $\eta$, perpendicular to the incident x-ray beam by 3.2\,deg.

\begin{figure}[tbh]
\includegraphics[width=0.8\linewidth]{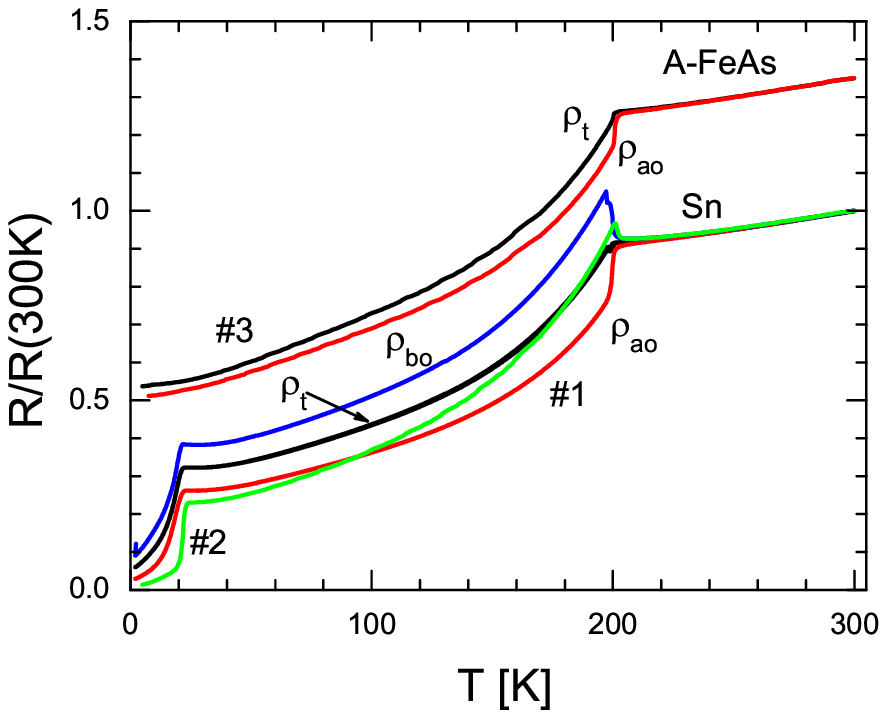}
\includegraphics[width=0.8\linewidth]{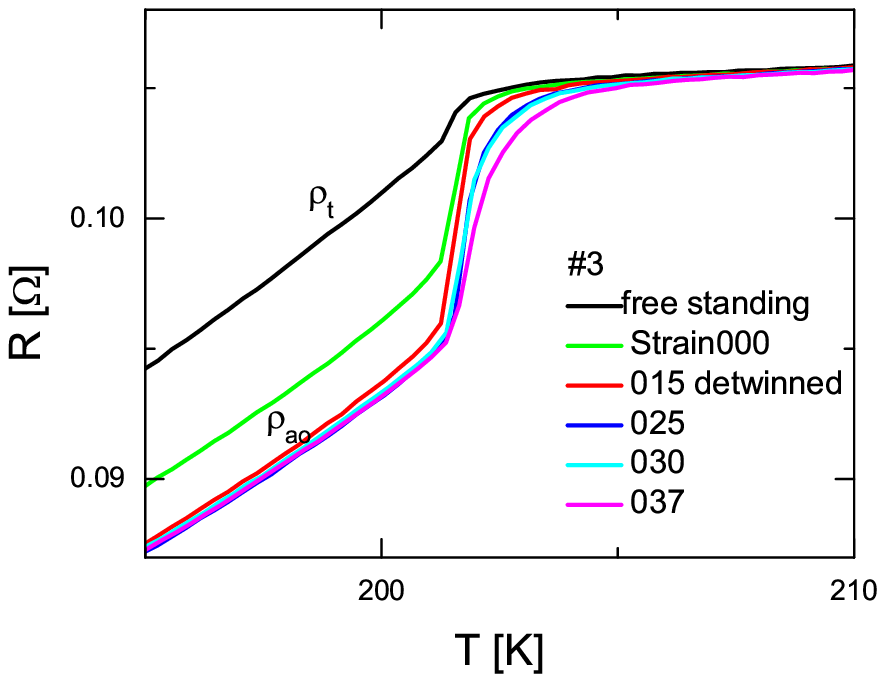}
\caption{ Top panel. Temperature dependence of resistivity measured along the tetragonal [110]$_T$ direction in twinned ($\rho_t$, black curves) and strain-detwinned ($\rho_{ao}$, red curves) states for sample \#1 grown from Sn flux and sample \#3 grown from FeAs flux with subsequent annealing (A-FeAs).
Blue line shows calculated temperature-dependent resistivity for \#1 in the direction transverse to the strain, $\rho_{bo}\equiv 2 \times \rho_t -\rho _{ao}$ \cite{detwinning}. Green line shows temperature-dependent resistivity for another Sn-grown sample \#2, partially detwinned by application of stress with preferable orientation of domains in the $b$ orthorhombic direction. The bottom panel shows a zoom of the temperature dependent resistivity for sample \#3 in the vicinity of the structural transition  as a function of relative displacement of the horseshoe sides (in arbitrary but monotonically increasing units). The free standing sample was measured before being mounted on the horseshoe. Fixing the sample to the horseshoe creates some strain even without any additional displacement from the push-screw (Strain 000) and partially detwins it. The pressure value for the highest displacement was estimated as in MPa range. Red curve (015 displacement) shows resistivity in the detwinned state, showing a sharp transition with no features above $T_{\textrm{TO}}$. With further increase of strain the feature at the transition broadens and reveals strain-induced resistivity anisotropy in nominally tetragonal phase. }
\label{resistivity}
\end{figure}

\section{Results}


To obtain twin-free regions in the samples, crystals were strained at room temperature and kept under strain while measuring temperature-dependent resistivity and studying domain images with polarized
light microscopy at 5 K. Strain was progressively increased until no twins were observed in the area between potential contacts. The images in the bottom panels of Fig.~\ref{detwinning}  illustrate
the effect of the strain in detwinning Sr122, with a zoom of the spots on the sample (as shown in the top panel)  in the strained (left) and unstrained (right) areas. The entire area of the crystal between the straining contacts ($\sim 1.8 \times 0.6$ mm$^2$) was found to be essentially free of twins under polarized microscopy.

Scanning the x-ray beam across the sample allows a spatially resolved characterization of the domain population as demonstrated in the second panel of Fig.~\ref{detwinning}. The map shows spatial distribution of the percent volume fraction of the crystallographic domains with distortion along the strain in the crystal. This was obtained by the analysis of the x-ray intensity distribution in the 6~K pattern arising from splitting (220)$_T$ peak, as schematically shown in the third  panel in Fig.~\ref{detwinning}. In the twinned area of the crystal, four possible crystallographic domains, with orientations along $a_o$ (O2 and O4) and $b_o$ (O1 and O3) are equally populated, leading to four equal intensity spots. Application of strain makes formation of domains with the $a_o$ axis along the strain energetically favorable, therefore in the strained regions, only two reflections are visible. The integrated intensities of these two reflections reveal relative orthorhombic domain populations of about 96\% and 4 \% for the two visible reflections.

As can be seen from the map, areas under the soldered contacts show random domain populations. In the strained part of the crystal between potential contacts, the volume fraction of a single domain reaches above 90\%. Since we do not see any other domains in the polarized microscopy image, we come to the conclusion that the domain population may have depth profile. This would be naturally expected in our experiment, in which deformation is applied through the contacts soldered to one sample surface and can lead to depth profile in strain distribution. Despite the fact that detwinning is not complete in the bulk, we are able to get a clear trend in the temperature- dependent resistivity, since contact resistance is much smaller than sample resistance and thus contacts work to shorten the unstrained areas. As can be seen from the domain distribution map (second from top panel in Fig.~\ref{detwinning}), current flow between potential contacts (excluding the area of the contacts themselves, shown with dashed line) proceeds through the area in which 3 out of 24 pixels have a preferred domain population of 63 to 67\%, 3 more pixels have a domain population of in 75 to 78\% range, while the other 17 pixels have a volume population above 80\% and block any direct current path between the contacts through the areas with low-percent domain population.

\begin{figure}[tb]
\includegraphics[width=0.7\linewidth]{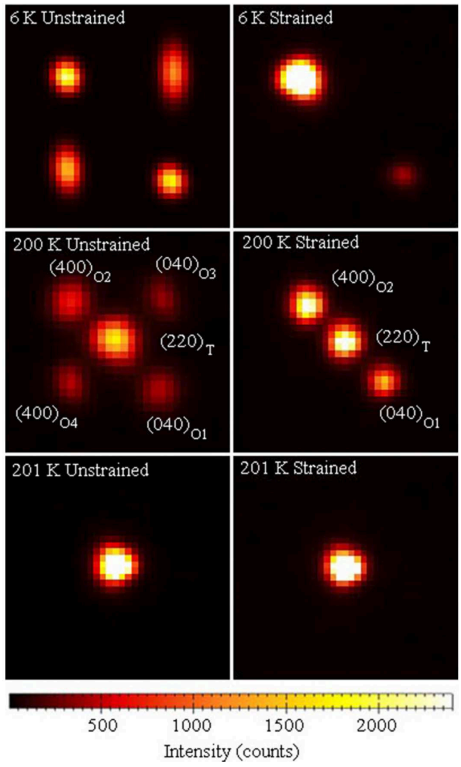}
\caption{ Temperature evolution of the (220)$_T$ peak (in the tetragonal phase crystallographic notation) of high energy x-ray Laue patterns in the  strain-free (twinned, left column of images) and  strained (right column of images) parts of the crystal, as shown in Fig.~\ref{detwinning}. At 6~K, the base temperature of our experiment (top panel), four peaks in the twinned part of the crystal correspond to four equivalent crystallographic directions of distortion in four equivalent domains, with very close to equal populations ranging between 23 and 26\% of the full integrated peak intensity. In the strained portion of the crystal, the intensity is distributed very unevenly, with the dominant spot comprising $\sim$96\% of the integrated peak intensity, the second spot approximately 4\%, while the other two peaks go below our resolution limit ($\sim$ 0.1\%). The orthorhombic peaks are observed all the way to the temperature of the structural, orthorhombic to tetragonal, phase transition. Phase coexistence of the orthorhombic and tetragonal peaks
at 200 K (second panel) clearly illustrates that (1)  the structural transition remains at the same temperature and is first order; (2) the domain population clearly changes with temperature.
The phase coexistence disappears abruptly within a 1~K step, as seen in both the strained and unstrained regions of the sample (bottom panels) in 201~K image, where the orthorhombic peaks are completely gone.
 }
\label{peaks}
\end{figure}

Resistivity measurements were performed using the same contacts on samples before and after application of strain. Nearly complete detwinning of the crystal, leads to a notable change of the temperature dependence of its resistivity. In Fig. ~\ref{resistivity} we show the resistivity of the same crystals \#1 and \#3 measured in the twinned and de-twinned states. The partial superconductivity in Sn-grown samples at 20~K is due to surface strain \cite{Saha} associated with cleaving and shaping the sample and is not focus of this study. This trace superconductivity is not observed in the annealed samples. The resistivity, $\rho (T)$, of unstrained samples cut along the [110]$_T$ direction is very close to that measured on samples from the same batch with current along the  [100]$_T$ direction \cite{anisotropypure}. It shows a feature at the structural/magnetic transition at $\sim$202~K. Straining the crystal gradually increases its resistivity at 300~K, however, the use of 0.05 mm diameter wires notably reduced fatigue, as compared with $\sim$1\% value per strain as found in Ca122 and Ba122 compounds in our initial study \cite{detwinning}.

\begin{figure}[tb]
\includegraphics[width=0.8\linewidth]{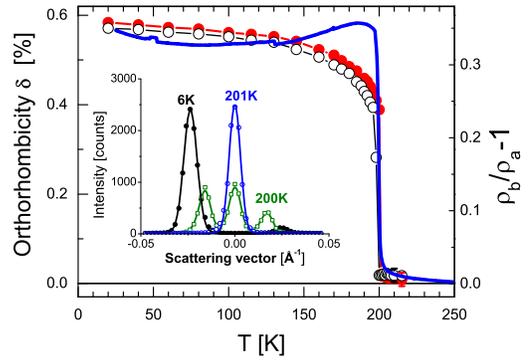}
\caption{ Temperature dependence of the structural order parameter $\delta=\frac{(a_o-b_o)}{(a_o+b_o)}$ as determined from the analysis of the (220)$_T$ spot splitting in single domain and twinned areas of the SrFe$_2$As$_2$ single crystal. The positions of the peaks were determined from the fit of the pixel profile to a Gaussian, as shown in the inset for strained area at three characteristic temperatures. Solid line shows anisotropy of electrical resistivity (right scale), normalized to match the magnitude of the structural order parameter at low and high temperature.
Anisotropy peaks below the transition and finds a very small residual value above the transition, coinciding within error bars with the magnitude of the strained-induced lattice distortion.
 }
\label{order_parameter}
\end{figure}

Temperature dependent resistivity, measured with the current along the strain direction, $\rho_{ao}(T)$, shows a sharp drop (17\% resistivity decrease in less than 1K change in \# 1) immediately below the transition temperature
$T_{\textrm{TO}}$, as opposed to a mild slope change in the twinned crystals. [This sharp drop, as corroborated by imaging in polarized light, is the main resistive signature of the detwinned samples.]  Of note, the sharp feature at the transition remains at the same temperature, though the $\rho(T)$ dependence changes dramatically.

Since the resistivity of twinned crystals can be viewed as an average over four domain orientations associated with two in-plane crystallographic directions \cite{detwinning}, we can get an insight into the behavior of $\rho_{bo}(T)$ assuming that $\rho_{t}=[ \rho_{ao}(T)+\rho_{bo}(T)]/2$ \cite{detwinning}. Thus calculated $\rho_{bo}(T)$ for sample \#1 is also shown in the Fig.~\ref{resistivity}, and it suggests an increase of $\rho_{bo}$ below the transition.
A similar increase of the resistivity in the direction transverse to the strain and a decrease of the resistivity for the direction along the strain is observed in both Ba122 and Ca122. Moreover, the anisotropy value immediately below $T_{\textrm{TO}}$ is of similar magnitude to 1.2 (Ca122) and 1.5 (Ba122), see Fig.~\ref{comparison} below. Thus calculated $\rho_{bo}$ in pure Ba122 crystals is very close to actually measured in crystals with stress-detwinning \cite{Fisher}.

To check if the temperature dependence of $\rho_{bo}(T)$ is a good approximation to real behavior, we measured resistivity in a Sn-grown sample \#2 (shown in the top panel of Fig.~\ref{resistivity}), that was squeezed by applying mild stress through potential contacts. Squeezing leads to a preferential domain orientation with the short orthorhombic $b_o$ direction along the current path. Although the state thus obtained was not as fully detwinned, it revealed the expected trend in $\rho_{bo}(T)$ with increase at $T_{\textrm{TO}}$.

The resistivity of the sample \#1, used in the x-ray study, reveals very weak anisotropy above the transition. To check if this anisotropy is associated with intrinsic anisotropy of the unstrained state (i.e. nematicity) or induced by the strain vector itself breaking rotational symmetry, we performed a systematic study of resistivity as a function of applied strain on yet another sample, \#3. In the bottom panel of Fig.~\ref{resistivity} we show temperature-dependent resistivity of the sample mechanically detwinned with systematically increasing strain. Strain value can be estimated to be in the 1 to 5 MPa range. All curves were measured in identical thermal cycle conditions on warming at a rate of 1 K/min. As can be seen, the resistivity change above the transition in the samples strained barely enough to achieve a detwinned state (as found in polarized optical microscopy study) does not find any trace of anisotropy above the transition. With further strain increase, the resistivity changes its temperature dependence in the nominally tetragonal crystallographic phase and the transition is preceded by a range of decreased resistivity extending approximately 10~K above the transition. This observation suggests that small $\sim$1\% anisotropy found in sample \#1 above the transition is induced by strain vector.


In Fig.~\ref{peaks} we show the temperature evolution of the (220)$_T$ peak in x-ray Laue patterns obtained on sample \#1 of Sn-grown Sr122. Similar images were taken every 10~K at temperatures in the range up to 240~K with finer 1~K steps in the vicinity of the transition.
As can be seen in the images taken at 200~K (middle panels in Fig.~\ref{peaks}), the orthorhombic and tetragonal phases coexist at the transition, clearly showing a first order type of transition, as in previous studies in twinned samples \cite{Yan,Luton}. The magnitude of the orthorhombic distortion, seen as peak splitting, right below the transition is approximately 60\% of the distortion at 5~K. At 201~K only the tetragonal peak is observed in both the strained and unstrained areas. These two observations clearly show that strain neither changes the first order character of the transition, nor its temperature $T_{\textrm{TO}}$ (the latter is consistent with the effect seen in resistivity measurements).


To get further insight into the behavior of the structural order parameter, $\delta \equiv \frac{(a_o-b_o)}{(a_o+b_o)}$, we made a quantitative analysis of the temperature-dependent x-ray peaks (shown for selected temperatures in Fig.~\ref{peaks}). The peak position was determined by fitting the intensity to a Gaussian shape, as shown in inset in Fig.~\ref{order_parameter}.

The temperature dependence of the order parameter for the orthorhombic phase in strained and strain-free parts of the sample is shown in Fig.~\ref{order_parameter}. The difference between the curves for strained and unstrained parts of the crystal as well as the tiny residual orthorhombicity above $T_{\textrm{TO}}$ reflect the residual effect of strain.
For comparison in Fig.~\ref{order_parameter} we show the calculated resistivity anisotropy of sample \#1 in the orthorhombic  plane, $\rho_{bo}/\rho_{ao}-1$. The value of $\rho_{ao}$ was measured directly in detwinned state of the sample. The value of $\rho_{bo}$ was calculated from $\rho_{ao}$ and resistivity measured in a twinned state of the sample, $\rho _t$, assuming random statistic averaging. As can be seen from comparison of the two quantities in Fig.~\ref{order_parameter}, their relative changes above the transition are coinciding within error bars. Together with systematic evolution of resistivity in the tetragonal phase as a function of strain, Fig.~\ref{resistivity}, this observation suggests that the tiny effect in resistivity above the transition comes from permanently applied strain.
In our high resolution and high dynamic range x-ray measurements we can exclude the contribution of local orthorhombic areas \cite{Vaknin} in the tetragonal phase at the level of approximately 0.1\% volume.

\begin{figure}[tb]
\includegraphics[width=0.8 \linewidth]{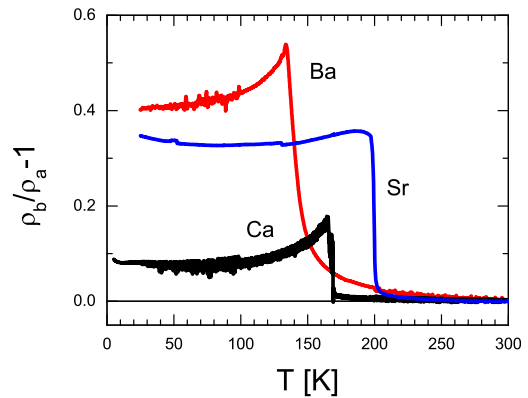}
\caption{ Temperature dependence of resistivity anisotropy, $\rho_{bo}/\rho_{ao}-1$ in three AFe$_2$As$_2$ compounds. The anisotropy monotonically increases with ionic radius of the rare earth element, peaks at or slightly below the structural transition, and then remains relatively constant. Notable anisotropy above the transition is observed only in A=Ba compound, with a weakly first order character of the structural transition.
 }
\label{comparison}
\end{figure}


In Fig.~\ref{comparison} we plot resistivity anisotropy as determined from measurements on three parent compounds of 122 iron pnictide superconductors. In all three materials the anisotropy takes its maximum value at or slightly below $T_{\textrm{TO}}$, then decreases on further cooling and becomes constant below approximately $T_{\textrm{TO}}/2$. This temperature dependence is in anti-correlation with both the degree of orthorhombicity and the long-range magnetic moment, which monotonically increase with cooling below $T_{\textrm{TO}}$. Since the magnetic order parameter develops more gradually \cite{Vaknin}, this feature may be associated with nematic fluctuations of the magnetic order below the strongly first-order structural transition.
The magnitude of the resistivity anisotropy above the tetragonal-to-orthorhombic structural transition anticorrelates with the sharpness of the first order transition. As suggested by our findings in samples with notably improved strain control, the anisotropy above the transition is induced by applied uniaxial strain. The anomalously large response to very small strain suggests that the crystals above the transition are very soft and responsive to strain.

\section{Conclusions}

In conclusion, electrical resistivity of SrFe$_2$As$_2$ in the orthorhombic phase reveals unusual electronic anisotropy with a resistivity decrease along the $a$-axis (direction of antiferromagnetic spin ordering) and increase along the $b$-axis (ferromagnetic chain direction). This behavior and temperature dependence of the anisotropy, $\rho_{bo}(T)/\rho_{ao}(T)$, with a maximum at/or slightly below $T_{\textrm{TO}}$, is similar in all parent AFe$_2$As$_2$ compounds \cite{detwinning}. The magnitude of the anisotropy monotonically increases with the ionic radius of the alkaline earth element, A. The resistivity anisotropy in SrFe$_2$As$_2$ is negligible above the strongly first order structural transition. However, the anisotropy in the tetragonal phase is easily induced by the application of a mild strain in the MPa range, suggesting a strong responsiveness of the compound. Our results suggest that electronic nematicity may not be observed in materials with a strong first order character of the structural transition. Additional studies are needed to clarify the effect of the strain itself on the electronic anisotropy in BaFe$_2$As$_2$ based materials to study the origins of the experimentally observed electronic anisotropy (nematicity) above the structural transition.

\subsection{Acknowledgement}

We thank D. Robinson for the excellent support of the high--energy x-ray scattering study. Use of the Advanced Photon Source was supported by the U. S. Department of Energy, Office of Science, under Contract No. DE-AC02-06CH11357. Work at the Ames Laboratory was supported by the U.S. Department of Energy, Office of Basic Energy Sciences, Division of Materials Sciences and Engineering under contract No. DE-AC02-07CH11358. R.H. acknowledges support from AFOSR-MURI grant \#FA9550-09-1-0603. R. P. acknowledges support from Alfred P. Sloan Foundation.



\begin{thebibliography}{58}



\bibitem{Rotter}M. Rotter, M. Tegel, and D. Johrendt, Phys. Rev. Lett. \textbf{101}, 107006 (2008).

\bibitem{Yan} J.-Q. Yan, A. Kreyssig, S. Nandi, N. Ni, S. L. Bud?ko, A. Kracher, R. J. McQueeney, R. W. McCallum, T. A. Lograsso, A. I. Goldman, and P. C. Canfield
Phys. Rev. B \textbf{78}, 024516 (2008).


\bibitem{Ca-phasetransition} N. Ni, S.~Nandi, A.~Kreyssig, A.~I.~Goldman, E.~D.~Mun, S.~L.Bud'ko, and P.~C.~Canfield,
Phys. Rev. B \textbf{78},
014523 (2008).



\bibitem{Milton} M. S. Torikachvili, S. L. Bud'ko, N. Ni and P.C. Canfield, Phys. Rev. Lett. \textbf{101}, 057006 (2008).


\bibitem{domains} M. A. Tanatar, A. Kreyssig, S. Nandi, N. Ni, S. L. Bud'ko, P. C. Canfield, A. I. Goldman, and R. Prozorov,
Phys. Rev.B \textbf{79}, 180508 (R) (2009).

\bibitem{detwinning} M. A. Tanatar, E. C. Blomberg, A. Kreyssig, M. G. Kim, N. Ni, A. Thaler, S. L. Bud\textquoteright{}ko, P. C. Canfield, A. I. Goldman, I. I. Mazin and R. Prozorov,
Phys. Rev. B \textbf{81}, 184508 (2010).


\bibitem{Fisher} Jiun-Haw Chu, J. G. Analytis, K. De Greve, P. L. McMahon, Z. Islam, Y. Yamamoto, and I. R. Fisher,
Science {\bf 329}, 824 (2010).

\bibitem{Eu} J. J. Ying, X. F. Wang, T. Wu, Z. J. Xiang, R. H. Liu, Y. J. Yan, A. F. Wang, M. Zhang, G. J. Ye, P. Cheng, J. P. Hu, X. H. Chen, arxiv 1012.2731

\bibitem{RotterSDW}
M. Rotter, M. Tegel, D. Johrendt, I. Schellenberg, W. Hermes, R. Poettgen
Phys. Rev. B {\bf 78}, 020503 (2008).

\bibitem{NiNiCo} N. Ni, M. E. Tillman, J.-Q. Yan, A. Kracher, S. T. Hannahs, S. L. Bud'ko, and P. C. Canfield,
Phys. Rev. B \textbf{78}
214515 (2008).




\bibitem{Fradkin}
E. Fradkin, S. A. Kivelson, M. J. Lawler, J. P. Eisenstein, A. P. Mackenzie, Ann. Rev. Cond. Mat. Phys. \textbf{1}, 153 (2010).
arxiv:0910.4166
%

\bibitem{Johannes}I.~I.~Mazin, and M.~D.~Johannes, Nature Phys.
\textbf{5}, 141 (2009). 



\bibitem{327} R. A. Borzi, S . A. Grigera, J. Farrell, R. S. Perry, S. J. S. Lister, S. L. Lee, D. A. Tennant, Y. Maeno, and  A. P. Mackenzie,
Science {\bf 315}, 214 (2007).



\bibitem{quantumHall}M. P. Lilly, K. B. Cooper, J. P. Eisenstein, L. N. Pfeiffer, K. W. West, Phys. Rev. Lett. {\bf 82}, 394 (1999).

\bibitem{nematiccuprates}  V. Hinkov, D. Haug, B. Fauqué, P. Bourges, Y. Sidis, A. Ivanov, C. Bernhard, C. T. Lin, and B. Keimer
Science {\bf 319}, 597 (2008).


Y. Ando, S. Segawa, S. Komiya, and A.N. Lavrov, Phys. Rev. Lett. {\bf 88}, 137005 (2002).




R. Daou, J. Chang, David LeBoeuf, Olivier Cyr-Choinière, Francis Laliberté, Nicolas Doiron-Leyraud, B. J. Ramshaw, Ruixing Liang, D. A. Bonn, W. N. Hardy and  Louis Taillefer
Nature {\bf 463}, 519 (2010).




\bibitem{Vaknin}
Y. Lee, D. Vaknin, H. F.  Li, W. Tian, J. L.  Zarestky, N. N, S. L. Bud'ko, P. C. Canfield, R. J. McQueeney, and B. N. Harmon,
Phys. Rev. B \textbf{81}, 060406 (2010).





\bibitem{Suchitra} J. Gillet, S. D. Das, P. Syers, A.K.T. Ming, J.J. Espresso, C. M. Petrone, and S.E. Sebastian, arxiv 1005.1330


\bibitem{contacts_SUST}
M. A. Tanatar, N. Ni, S. L. Bud'ko, P. C. Canfield,and R. Prozorov,
Supercond. Sci. Technol. {\bf 23}, 054002 (2010).


\bibitem{Kreyssig07}
A.~Kreyssig, S.~Chang, Y.~Janssen, J.\,W.~Kim, S.~Nandi, J.\,Q.~Yan, L.~Tan, R.\,J.~McQueeney, P.\,C.~Canfield, and A.\,I.~Goldman,
Phys. Rev. B \textbf{76}, 054421 (2007).



\bibitem{domains2}
R. Prozorov, M. A. Tanatar, N. Ni, A. Kreyssig, S. Nandi, S. L. Bud'ko, A. I. Goldman, and P. C. Canfield,
Phys. Rev.
B \textbf{80}, 174517 (2009).



\bibitem{Saha}S. R. Saha, N. P. Butch, K. Kirshenbaum, and J. Paglione,
Phys. Rev. Lett. \textbf{103}, 037005 (2009).

\bibitem{anisotropypure}  M. A. Tanatar, N. Ni, G. D. Samolyuk, S. L. Bud'ko, P. C. Canfield, and R. Prozorov,
Phys. Rev. B \textbf{79},
134528 (2009).


\bibitem{Luton}
J.C. Loudon, C. J. Bowell, J. Gillett, S. Sebastian, and P.A.Midgley,
Phys. Rev. B \textbf{81}, 214111 (2010).


\end{thebibliography}
\end{document}